# Electrostimulation of Brain Deep Structures in Parkinson's Disease.

**Elcin Huseyn**
Azerbaijan State Oil and Industry
University, Baku, Azerbaijan
elcin.huseyn@asoiu.edu.az

**Abstract:** The study involved 56 patients with advanced and late stages of Parkinson's disease, which could be considered as potentially requiring neurosurgical treatment - electrical stimulation of deep brain structures. An algorithm has been developed for selecting patients with advanced and late stages of Parkinson's disease for neurological treatment-implantation of a system for electrical stimulation of deep brain structures in distant neurosurgical centers, which includes two stages for patients with limited mobility - outpatient and inpatient. The development of an algorithm for referral to neurological treatment has shortened the "path" of a patient with limited mobility from a polyclinic to a neurological center. Electro stimulation of deep brain structures in Parkinson's disease significantly improved the condition of patients - to increase functional activity by 55%, reduce the severity of motor disorders by 55%, and reduce the dose of levodopa drugs by half.

## 1. Introduction

In recent years, significant progress has been made in the drug therapy of Parkinson's disease. However, drug therapy for advanced and late stages is not always effective and is sometimes accompanied by the occurrence of side effects in the form of disabling symptoms that significantly limit the life of patients [1, 2].

The greatest limitation of life is caused by motor fluctuations - fluctuations in the patient's motor function up to complete immobility. The development of motor fluctuations depends on the duration of the disease. They occur in 30-60% of patients with a disease duration of 4-5 years, and with a disease duration of 15 years - in 96% of patients. Medicinal dyskinesias occur in 50-90% of patients receiving levodopa drugs for a long time (more than 5 years) [1, 2].

Thus, the treatment of Parkinson's disease in the advanced and later stages is a difficult task, requires significant economic costs, both, direct, associated with the need to prescribe many drugs, and indirect, associated with the restriction of the work activity of the patient and the caregivers [3].

Electro stimulation of deep brain structures is recognized as a highly effective method of treating late and advanced stages of Parkinson's disease, overcoming motor fluctuations and drug dyskinesias [4, 5].

The advantage of electrical stimulation of the deep structures of the brain, in comparison with destructive methods of treatment, is the ability to obtain the maximum effect with minimal side effects [6, 7].

In the case of successful electrical stimulation of the subthalamic nucleus, the severity of the shutdown periods decreases, their duration is reduced; the dose of antiparkinsonian drugs can be reduced by 50% [8, 9].

However, there is a fairly large number of contraindications and restrictions for electrical stimulation of deep brain structures in Parkinson's disease [4, 10].

If these limitations and contraindications are underestimated, the erroneous referral to a specialized center for patients with limited mobility aggravates the disorders of their neuropsychic status.

Incorrect selection of patients referred to specialized centers at the preliminary stage is over 60%, which leads to refusal in neurosurgical treatment or is one of the main reasons for the ineffectiveness of this type of treatment [11, 12].

It is obvious that careful selection of patients is needed for surgical intervention - stimulation of deep brain structures in distant neurosurgical centers [13, 14].

The aim of this work was to develop an algorithm for selecting patients with Parkinson's disease for neurological treatment - implantation of a system for electrical stimulation of deep brain structures.

## 2. Materials and research methods.

We examined 56 patients with advanced and advanced stages of Parkinson's disease, which could be considered as potentially requiring neurosurgical treatment. Among them, 23 men and 33 women aged from 29 to 70 years, the average age was 57.63 ± 7.5 years, the age of debut - 49.1.1 ± 7.4 years, the average duration of the disease - 8.64 ± 3, 29 years old. Parkinson's disease of the 3rd severity according to Hen-year was observed in 40 people, the fourth - in 16 patients. Disability in the 1st group was in 2 patients, the 2nd group - in 34 patients, the 3rd group - in 17 people; only 6 patients who reached retirement age did not have a disability. All patients underwent an in-depth neurological examination, which included the use of the unified Parkinson's Disease Scale in the "on and off" periods, the study of cognitive functions using a short scale for assessing mental status, and magnetic resonance imaging of the brain (MRI). Optimization of drug therapy was carried out using metabolic precursors of dopamine (levodopa-benserazide, levodopa-carbidopa), drugs-stimulators of dopamine postsynaptic receptors (dopamine receptor agonists - pronoran (piribedil), pramipexole, ropinilyaerine-catabolism), drugs, inhibitor -razagiline), blockers of glutamate NMDA receptors that inhibit reuptake, increase release, mild stimulation of dopamine receptors (amantadines), central blockers of M-cholinergic receptors (trihexyphenidyl, bipyridine). If it was impossible to select therapy at the outpatient stage, patients were admitted to a neurological hospital for further selection of therapy and selective for stimulation of deep brain structures.

## 3. Results and their discussion.

At the first stage, at the polyclinic level, patients with advanced and late stages of Parkinson's disease were selected, with a disease duration of more than five years, who had disabling motor symptoms, despite the use of high doses of levodopa at the age of up to 70 years, which could be considered as candidates for stimulation deep structures of the brain. 56 people were initially selected. All patients selected for referral to neurological treatment had a degree of functional activity, according to the Schwab – England scale, on average, 25 ± 4.4%. The duration of the "on" period was 1–1. 5 hours, or the patients did not notice the effect of the use of levodopa drugs (3 patients). Among the disabling motor symptoms were the phenomenon of wear of the end of the dose with the development of "on-off" symptoms, insufficient switching on, delayed switching on, unpredictable "switching off", lack of switching on. In addition, end-dose dyskinesias were observed in the form of painful dystonia in 14 patients, peak dose dyskinesias in 12 patients, and biphasic dyskinesias in 4 patients.

To determine the response of patients to levodopa drugs, a selection of drug therapy was made with subsequent monitoring of its effectiveness. The results of an MRI of the brain, the quality of tests for cognitive functions, psychoemotional and somatic status were assessed.

The most important goal of these studies at the outpatient stage was to identify contraindications for neurological treatment in the form of electrical stimulation of the deep brain structures.

There were three groups of contraindications. The first group was associated with the characteristics of the course of Parkinson's disease in a particular patient. This group of symptoms and conditions included: insufficient response to levodopa, in which surgery is ineffective; cognitive dysfunction; untreated depression; patient disability due to "nondopamine", mainly axial symptoms - gait disturbances, camptocormia, gross postural instability, dysphonia. The second group of contraindications included concomitant diseases, which were contraindications to implantation of a system for electrical stimulation of deep brain structures, uncontrolled arterial hypertension, concomitant cerebrovascular pathology with multiple foci of discirculation in the brain, in which there is a high risk of postoperative intracranial hemorrhages. The third group of contraindications was social. This is reluctance or inability, including on the part of the patient's relatives, to cooperate during a surgical procedure or in the process of postoperative programming [13, 14].

In case of correcting factors, such as depression or arterial hypertension, after the therapy, patients were again considered as candidates for neurological treatment.

Special attention was required to carry out a differential diagnosis between Parkinson's disease and atypical parkinsonism, first, multisystem atrophy, in which the implantation of a system for stimulating deep brain structures and neurostimulation provokes the progression of the disease - these are patients with atypical parkinsonism syndromes, among which multisystem atrophy takes the leading place [15].

Alarming symptoms in such patients are early postural instability, early orthostatic hypotension, falls, levodoporesistance, early urinary disorders in the framework of autonomic failure.

Differential diagnosis of multisystem atrophy with late stages of Parkinson's disease in the presence of non-motor manifestations in the form of autonomic failure presented the greatest difficulties.

If necessary, complex cases were discussed online with neurosurgeons of the centers performing surgical interventions.

During the diagnostic measures carried out at the prehospital stage, 8 patients were identified who had contraindications to stimulation of the deep brain structures. Among them - an unsatisfactory response to levodopa drugs (less than 30% improvement in motor function) was detected in 2 patients, pronounced cognitive impairment in 4 patients (the number of points on a short scale for assessing mental status 18, 19, 20 and 22 points, respectively).

In two cases, there was marked postural instability with falls. In one case, the patient showed pronounced discirculatory changes around the basal ganglia according to MRI data. In one case, the patient showed clinical and neuroimaging signs of multisystem atrophy in the form of progressive autonomic failure, levodoporesistance, pyramidal insufficiency, cerebellar symptoms in combination with atrophy of the Pons, cerebellum, and middle peduncles of the cerebellum. When optimizing drug therapy in 14 patients with stage 3 Parkinson's disease, it was possible to achieve compensation for motor disorders due to the combined use of levodopa drugs, dopamine receptor antagonists, MAO-B inhibitor, amantadines, changes in the levodopathic regimen - increasing the dose of drugs, increasing the frequency of administration, using the combined preparation of levodopa-carbidopa - enttacapone - stalevo. A decrease in the severity of motor symptoms was also achieved in two patients with severe postural instability. All patients with cognitive impairments, along with the optimization of levadopathy, were prescribed akatinol memantine in an increasing dosage, followed by control of cognitive impairments. After correction of motor fluctuations, patients with postural instability were given recommendations to improve motor control — postural training, walking with two sticks, and the use of visual support. When levodoporesistance was detected, PK-mark was added to the treatment.

In 32 patients, it was not possible to achieve an improvement in motor function at the outpatient stage; on the other hand, they did not have any contraindications to neurological treatment at the prehospital stage. All these patients belonged to the group of patients with limited mobility, had significant disabilities, all patients had disabling motor symptoms of Parkinson's disease, which required correction in a hospital setting [16]. Thus, the preparation of patients for neurological treatment, in addition to outpatient treatment, included a hospital stage of selection.

At the hospital stage, the main therapeutic and diagnostic tasks were maximum optimization of drug therapy, including with the use of parenteral administration of drugs, with subsequent re-assessment of the motor part of the unified Parkinson's disease scale in the on-off periods, and assessment of the patient's functional capabilities using the Schwab-England scale. The total duration of the shutdown, diagnostics of all motor fluctuations and dyskinesias present in patients with video recording of syndromes, as well as detection of covert cardiovascular diseases, such as complex rhythm and conduction disturbances, detection of covert oncological diseases, were assessed. Targeted neuroimaging of the basal ganglia in patients with concomitant cerebrovascular pathology was carried out to identify foci of discirculation in the proposed implantation of electrodes. The benefits and limitations of deep brain stimulation were explained to patients and their families.

A total of 10 patients was selected for neurological treatment out of 32 patients who fully met the selection criteria.

In the remaining patients, the limiting circumstances were: for the first time diagnosed latent prostate cancer and basal cell carcinoma - 2 cases, cardiac arrhythmias and conduction disturbances requiring preliminary treatment by a cardiologist - 4 cases, uncontrolled arterial hypertension - 2 cases, depression - 2 cases, changes in the head brain in the area of the basal ganglia according to MRI data, which does not allow for effective neurosurgical intervention - 6 patients; multisystem atrophy - 2 patients; refusal of neurosurgical treatment - 4 patients.

When uncorrectable contraindications to the implantation of a system to stimulate the deep brain structures were identified, patients were optimized drug therapy in a hospital setting, and non-drug methods of treatment were also used, such as physical therapy, walking and speech training. Thus, the stationary stage of selection for patients with limited mobility with Parkinson's disease is a necessary part of the diagnostic algorithm and allows both to intensify drug therapy and to screen out patients with conditions that are unacceptable for neurosurgical treatment of Parkinson's disease.

Among the patients finally selected for neurosurgical treatment of Parkinson's disease, there were 5 men and 5 women aged 30 to 63 years, the average age was $55.8 \pm 11.8$ years. The age of onset of the disease ranged from 24 to 53 years, the average age of onset was $43.3 \pm 8.62$ years. The majority (7 patients) were patients with early-onset parkinsonism, with an onset age of up to 45 years. The duration of the disease ranged from 5 to 15 years. The average duration of the disease is $12.5 \pm 4.52$ years. Thus, in most patients, significant disabilities arose even before reaching retirement age.

In five cases, caring for the sick required the constant presence of the caregiver, which led to the need to restrict active work for the caregiver. All patients were diagnosed with advanced (third) or late (fourth) stages of Parkinson's disease. Motor fluctuations were detected in all subjects in the form of predictable and unpredictable shutdowns (10 people). Also, gait disorders were diagnosed in the form of a violation of the initiation of walking and "sticking" (8 people). Freezing was observed in 8 patients, painful dyskinesias of the end of the dose - in 7 people, gross biphasic dyskinesias - in 2 people, severe dyskinesias of the dose peak - in 1 patient. The greatest disability in patients selected for neurological treatment was caused by the "on-off" and "no on" phenomena. Moreover, if predictable shutdowns required modification of daily activity, unpredictable shutdowns, in addition to limiting life activity, led to panic reactions and provoked uncontrolled intake of levodopa drugs up to 8-10 tablets per day.

Neurosurgical treatment in the form of implantation of the ACTIVA PC system for deep electrical stimulation of the subthalamic nuclei on both sides was performed in 10 patients with Parkinson's disease at the N.N. Acad. N.N. Burdenko (Moscow) and at the Federal Neurosurgical Center (Tyumen).

After the neurological treatment, the mean score for the motor part of the unified Parkinson's disease scale decreased by 55% - from $83.4 \pm 14.1$ to $37.2 \pm 12.4\%$ ($p = 0.0001$). There was a decrease in dependence on others. According to the Schwab-England scale, daily activity increased from $24 \pm 5.2$ to $79 \pm 11\%$ ($p = 0.0001$). Four patients became completely independent from the others, partially in need of outside help - three patients, more frequent dependence on outside help was observed in three patients, however, even in these cases, the need for the constant presence of several caregivers disappeared.

In three cases, the caregivers were able to go to work, in one case the patient himself took over the responsibilities of caring for the child, one patient returned to work. Thus, because of the neurological treatment carried out, not only the medical, but also the social status of patients and caregivers has improved. The average dose of levodopa almost halved - from $1340 \pm 405$ to $697.5 \pm 256$ mg ($p = 0.0005$). In five patients with an initial duration of the "off" period of 51–75%, the "off" phenomenon was completely arrested in the postoperative period. For the remaining five people, with the same initial duration of the "off" period - 51–75% per day, it decreased to 25% per day or less ($p = 0.0001$).

The main problems remained gait disorders - in five patients, requiring the use of special techniques to improve the initiation of gait, postural instability - in 4 cases, speech disorders that interfere with contact - in 2 patients, which do not interfere with contact with others, but are obvious to the patient and his family - in 3 patients. Violations of speech and gait refer to nondopamine - dependent symptoms of Parkinson's disease, so the correction of such disorders was not the main goal of neurosurgical intervention. Reducing speech disorders was achieved by installing two programs that the patient switches independently - one for better motor functions, the other for improving speech. Gait disorders can be overcome by using non-drug treatments.

The most important thing for patients in the postoperative period was the disappearance of unpredictable shutdowns and a significant decrease in predicting shutdowns, as well as the disappearance of disabling dyskinesias. Thus, the quality of life of patients has significantly improved. Our results are consistent with the data of other researchers [7, 13, 17].

## 4. Conclusion

An algorithm has been developed for selecting patients with advanced and late stages of Parkinson's disease for neurological treatment - implantation of a system for electrical stimulation of deep brain structures in distant neurological centers. The algorithm for selecting candidates for a neurological treatment includes two stages for patients with limited mobility - outpatient and inpatient; at the outpatient stage, patients are selected according to general indications for neurosurgical treatment of Parkinson's disease; the stationary stage of selection made it possible to carry out the maximum compensation for the condition of the patients and to identify patients with latent somatic diseases. The development of an algorithm for referral to neurological treatment has shortened the "path" of a patient with limited mobility from a polyclinic to a neurological center. Electro stimulation of deep brain structures in advanced and late stages of Parkinson's disease significantly improved the condition of patients - to increase functional activity by 55%, reduce the severity of motor disorders by 55%, and reduce the dose of levodopa drugs by half.